# Parity-Dependent Moiré Superlattices in Graphene/*h*-BN Heterostructures: A Route to Mechanomutable Metamaterials

Wengen Ouyang, Oded Hod, and Michael Urbakh

*Department of Physical Chemistry, School of Chemistry and The Sackler Center for Computational Molecular and Materials Science, The Raymond and Beverly Sackler Faculty of Exact Sciences, Tel Aviv University, Tel Aviv 6997801, Israel.*
Email: odedhod@tauex.tau.ac.il

**Abstract**

The superlattice of alternating graphene/*h*-BN few-layered heterostructures is found to exhibit strong dependence on the parity of the number of layers within the stack. Odd-parity systems show a unique flamingo-like pattern, whereas their even-parity counterparts exhibit regular hexagonal or rectangular superlattices. When the alternating stack consists of seven layers or more, the flamingo pattern becomes favorable, regardless of parity. Notably, the out-of-plane corrugation of the system strongly depends on the shape of the superstructure resulting in significant parity dependence of its mechanical properties. The predicted phenomenon originates in an intricate competition between moiré patterns developing at the interface of consecutive layers. This mechanism is of general nature and is expected to occur in other alternating stacks of closely matched rigid layered materials as demonstrated for homogeneous alternating junctions of twisted graphene and *h*-BN. Our findings thus allow for the rational design of mechanomutable metamaterials based on Van der Waals heterostructures.

Heterojunctions of vertically assembled van der Waals (vdW) layered materials[1] have attracted great scientific and technological interest since they exhibit diverse physical properties and carry great technological potential in various fields including electronics, optics, mechanics, and tribology [1-7]. Among the numerous possible vdW heterostructures graphene and hexagonal boron nitride (*h*-BN) junctions have attracted particular interest [8] as potential building blocks for field effect transistors [9,10], thermoelectric devices [11], solar cells, LEDs, and photodetectors [12-14].

The versatility of these systems stems from their unique electronic [15-19], magnetic [20,21], mechanical [22], and frictional properties [23-25], which are tightly intertwined with the moiré superlattices naturally occurring in such heterostructures [26,27]. By varying the twist angle between two adjacent layers, the moiré superlattices can be modified in a controlled manner opening a route to tune the corresponding material properties [15-19,28,29]. Further control can be gained via changing the number of stacked layers within the heterojunction. This has been demonstrated for ABA-stacked few-layer graphene (FLG), whose electronic properties exhibit dependence on the number of layers parity (being either odd or even) [30,31]. To the best of our knowledge, no similar parity effect has been observed in the mechanical properties of layered materials heterostructures, to date.

In the present paper, we reveal a new phenomenon, where the symmetry of the moiré superlattice and the corresponding mechanical properties of vdW heterostructures depend on the parity of the number of layers. We demonstrate this for alternating few-layered graphene/*h*-BN stacks, where even layered structures present regular hexagonal or rectangular moiré patterns, whereas odd-layered structures exhibit a unique flamingo-like pattern with relatively small Poisson's ratio. Furthermore, the sign of Poisson's ratio depends on the loading direction thus presenting novel auxetic metamaterial behavior [32-34]. Notably, above a stack thickness of seven layers, our calculations indicate that the flamingo pattern becomes preferential for both even- and odd-layered stacks.

Our model system consists of an aligned (non-twisted) stack of alternating layers of graphene and *h*-BN, whose number of layers is varied between 2 to 16. Intralayer interactions are modeled using the REBO potential [35] and the Tersoff potential [36], for graphene and *h*-BN, respectively. Interlayer interactions are modeled using our recently developed anisotropic interlayer potential (ILP) [24,37-39]. To eliminate the edge effects, periodic boundary conditions are applied in the lateral directions (more details regarding the model system and the computational approach are provided in the Methods section). Due to the inherent intralayer lattice vectors mismatch of ~1.8% between graphene and *h*-BN layers, highly-corrugated moiré patterns with flat nearly optimally stacked regions separated by narrow elevated ridges of suboptimal staking [22,37,40], appear following geometry optimization.

In Fig. 1a the bilayer graphene/*h*-BN system is presented, showing the regular hexagonal moiré pattern with typical lateral dimensions of ~13.8 nm. Notably, when increasing the number of layers of the alternating stack, the competition between interlayer interactions at the different interfaces results in more involved moiré patterns (see Fig. 1b-e). The out-of-plane corrugation (peak-to-dip value) of these structures are summarized in Fig. 1f. Surprisingly, strong parity dependence is observed up to a stack thickness of 7 layers, where the out-of-plane corrugation of the odd numbered heterostructures is 4-5 Å larger than that of the even numbered counterparts. Above this thickness, the parity dependence disappears and the lowest energy moiré pattern we could identify strongly deviates from the regular hexagonal superstructure obtained in the bilayer case.

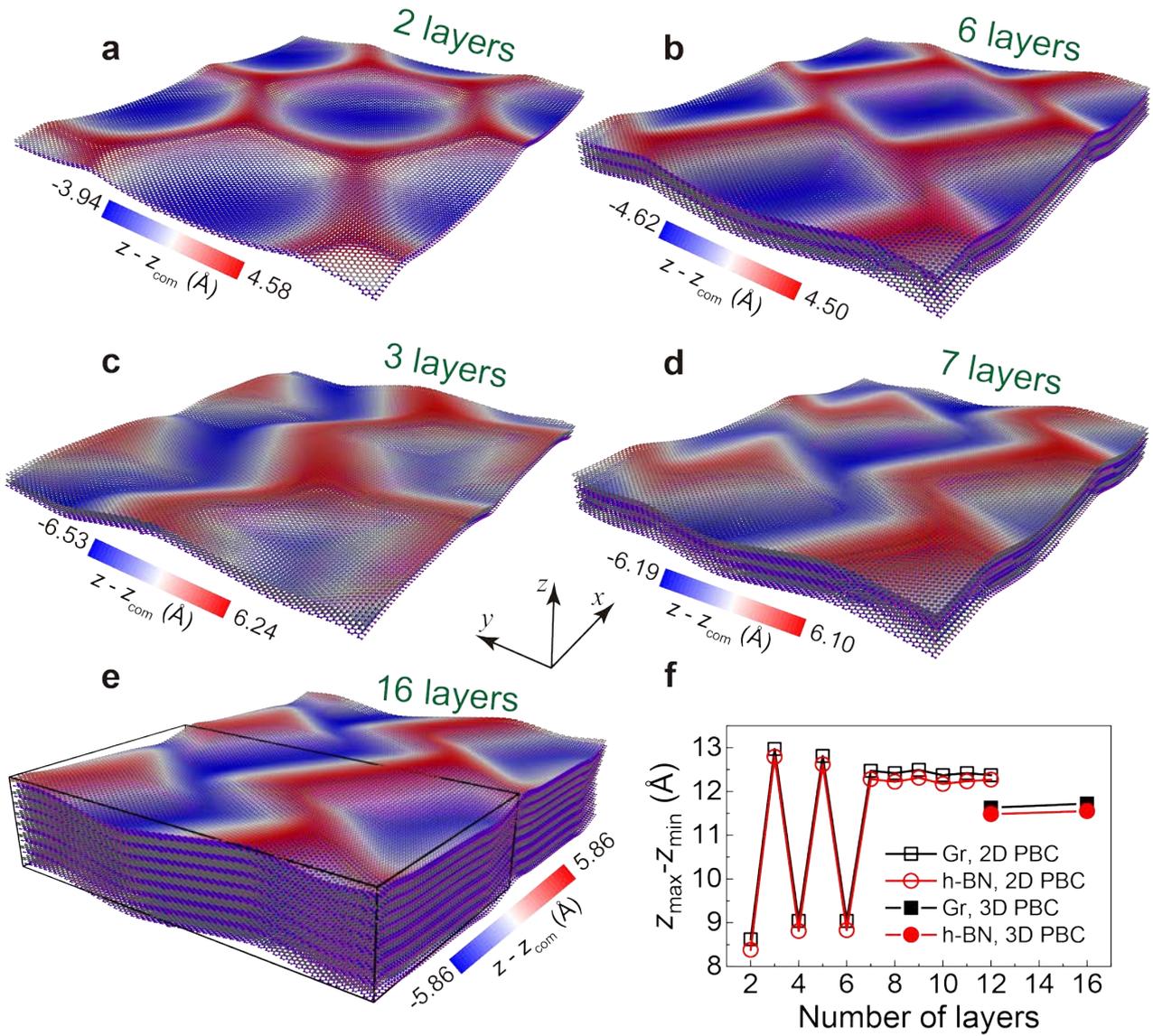

**Fig. 1. Optimized alternating graphene/*h*-BN heterostructures of various number of layers** obtained using two-dimensional (2D) periodic boundary conditions (PBC) in the lateral directions (**a**-**d**) and three-dimensional (3D) PBC (**e**). The atoms of the upper layers are colored according to their vertical height with respect to the center of mass of the layer (see corresponding color bars next to each panel). Mauve, blue, and gray spheres in the lower layers represent boron, nitrogen, and carbon atoms, respectively. For clarity, two unit-cells are presented (see the black box in panel **e** for the actual supercell dimensions used in the simulations). The out-of-plane corrugation (peak-to-dip value) in both graphene (black squares) and *h*-BN (red circles) layers as a function of the number of layers of the alternating graphene/*h*-BN heterostructures is presented in panel **f**. Open symbols represent calculations obtained using 2D PBC and full symbols represent calculations performed using 3D PBC with different super-cell thicknesses.

To understand the parity dependence of the deformation pattern, we plot in Fig. 2 the out-of-plane corrugation pattern for a consecutive series of alternating stacks of increasing thickness ranging from 2 to 9 layers. From this figure, the parity dependence is clearly evident, where the even stacks (up to 6 layers) present either the regular hexagonal structure or a related rectangular moiré structure, whereas the corresponding odd stacks exhibit a unique superstructure that resembles the shape of a flamingo (see Fig. 2). As mentioned above, for stack thickness exceeding 6 layers the flamingo pattern is found to be the most stable one (see Sec. 3 of the Supporting Information for thicker heterostructures results). This is further supported by our 3D PBC calculations that exhibit the same preference for the flamingo-like superstructure (see Fig. 1e). Importantly, for the even-numbered stacks, we were able to identify only a single stable structure, whereas for the odd-numbered systems many meta-stable structures were obtained, among which the flamingo-like structure was the most stable one (see Sec. 1 and Movies 1 and 2 in the Supporting Information for further details).

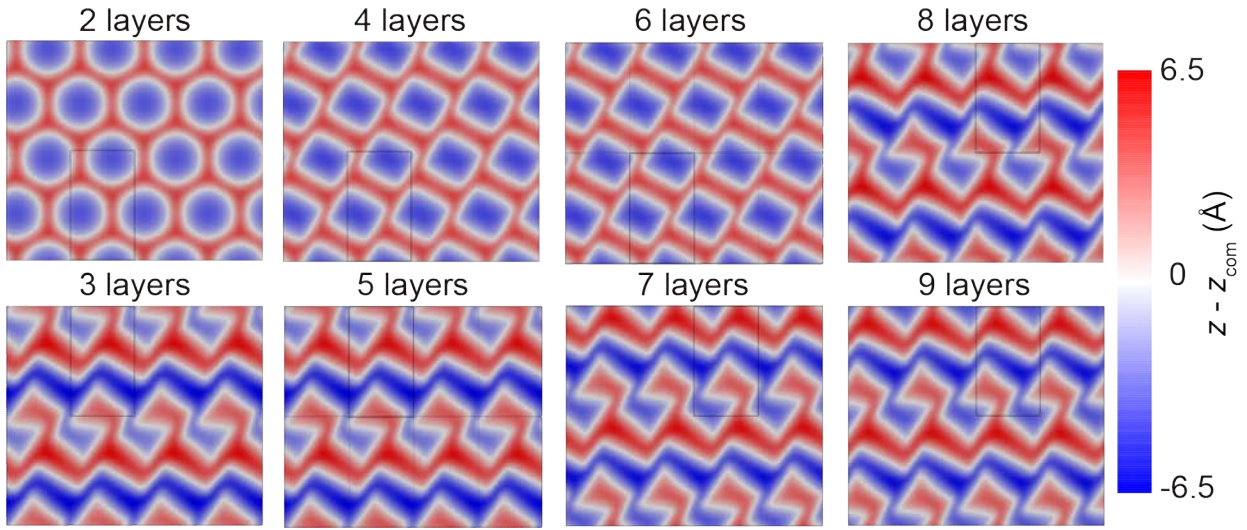

**Fig. 2. Parity dependence of the out-of-plane distortion superstructures for alternating graphene/$h$-BN heterostructures.** Top views of consecutive stacks of increasing thickness ranging from 2 to 9 layers are presented. The flamingo-shaped superstructure is clearly observed for the odd-stacks (bottom panels) and for the thicker even-stacks (top right panel). All results presented in this figure are obtained using 2D PBC boundary conditions optimizations. For clarity of the presentation, the simulation box (marked by a black rectangle in each panel) is multiplied in the lateral directions. The color scale denotes the atomic vertical height of the atoms in the top layer with respect to its center of mass.

The predicted parity dependence of the vertical deformation superstructure opens a route to design the mechanical properties of layered materials heterostructures. To demonstrate this, we evaluate Young's modulus and Poisson's ratio for graphene/$h$-BN alternating heterostructures of varying thickness (the corresponding simulation protocol is presented in the Methods section and raw simulation data can be found in Sec. 5 of the Supporting Information). The results are summarized in Fig. 3.

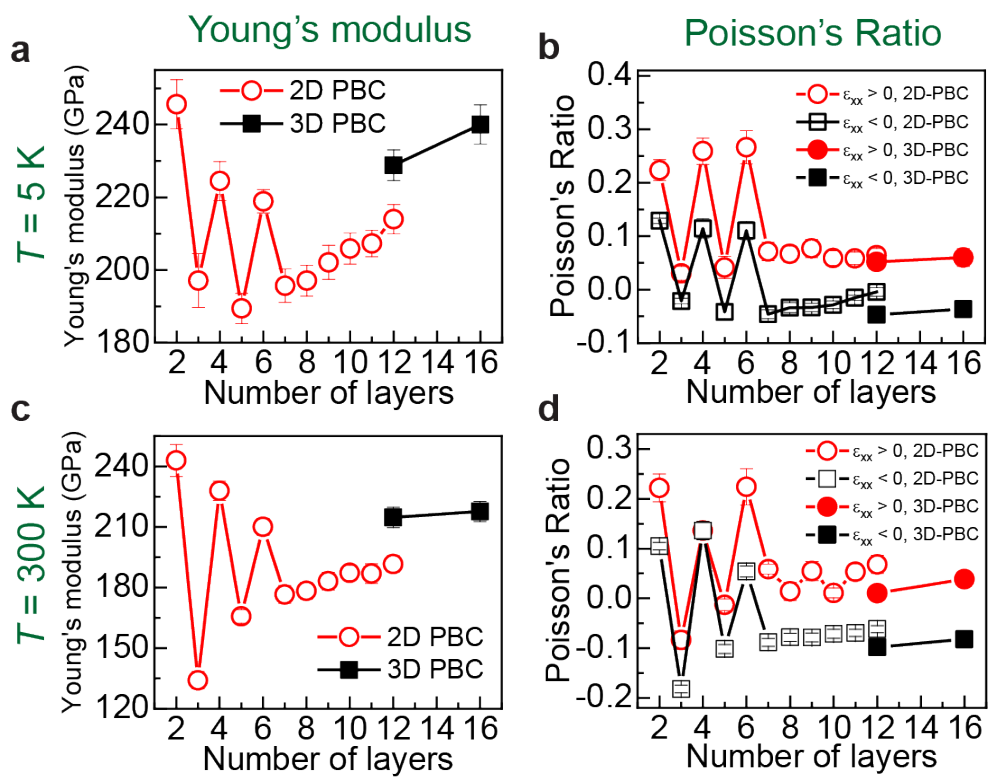

**Fig. 3. Parity dependence of the mechanical properties of alternating graphene/$h$-BN heterostructures.** (**a**, **c**) Young's modulus and (**b**, **d**) Poisson's ratio as functions of the number of layers evaluated at a temperature of $T = 5$ K (upper panels) and $T = 300$ K (lower panels), respectively. Open symbols (full symbols) represent calculations obtained using 2D (3D) PBC with different super-cell thicknesses. In panels **b** and **d**, results for both stretching (red circles) and compression (black squares) strains are presented to emphasize the dependence of Poisson's ratio on the loading direction. The error bars represent the standard deviation obtained in the calculation of the corresponding coefficients (see Figs. S8-S11 in the Supporting Information for further details).

From Fig. 3a-b it is evident that both Young's modulus and Poisson's ratio exhibit strong parity dependence (up to a factor of 2 and 50, respectively) when the number of layers is smaller than 7.

The parity dependence disappears for thicker stacks, consistent with the deformation pattern behavior discussed above. Increasing the temperature from 5 K to 300 K has minor effect on the parity dependence of the elastic constants (see Fig. 3c-d and Sec. 5 of the Supporting Information). The value of the Young's modulus for the alternating graphene/*h*-BN heterostructure (~240 GPa) is much lower than that of monolayer graphene (~1000 GPa) [41] and *h*-BN (~865 GPa) [42]. We attribute this to the highly corrugated nature of the heterostructures that can flatten upon loading while reducing in-plane covalent bond elongation stress (see Sec. 7 in the Supporting Information). For thin stacks ($N < 7$) at 5 K, the Poisson's ratio (Fig. 3b) of the even-numbered systems are positive and in the range of 0.1-0.3, whereas for odd numbered systems it is minute (~|0.04|) and changes sign when the direction of applied strain is varied. When the number of layers exceeds seven, the parity dependence of Poisson's ratio disappears and its value ranges between -0.046~-0.0045 and 0.058~0.077 for compression and stretching deformations, respectively (these small values should be considered qualitatively in light of the approximate nature of our classical molecular dynamics computational approach). The values of Poisson's ratio calculated at room temperature are very similar to those obtained at 5K (see Fig. 3d and Secs. 5 and 6 in the Supporting Information). We note that small Poisson's ratio materials have recently attracted much attention due to their potential utilization in many advanced applications [43-46].

The discovered parity dependence of the vertical deformation patterns is attributed to the intricate interplay between the large-scale moiré superlattices formed at each interface in the heterostructures. This suggests that the phenomenon is not limited to the case of graphene/*h*-BN heterostructures but should be observed for many alternating vdW layered stacks of closely matched lattice vectors, where large moiré superlattices are formed between adjacent layers. Notably, this can also be achieved in homogeneous alternating layered stacks, where adjacent layers assume a misaligned configuration. To demonstrate this, we constructed multilayer alternating twisted graphene and *h*-BN stack models, where the even numbered layers are rotated by 1.12° with respect to their odd numbered counterparts. As shown in Fig. 4, the out-of-plane deformation pattern of the twisted graphitic system exhibits clear parity dependence on the number of stacked layers with large variations of its out-of-plane corrugation. Similar results for *h*-BN are presented in Sec. 4 of the Supporting Information. This demonstrates the general nature of the predicted superstructure parity dependence phenomenon. Together with the nearly vanishing Poisson's ratio exhibited by the flamingo-shaped superstructure materials, our findings therefore suggest the possibility to construct a new type of metamaterials with tunable mechanical properties.

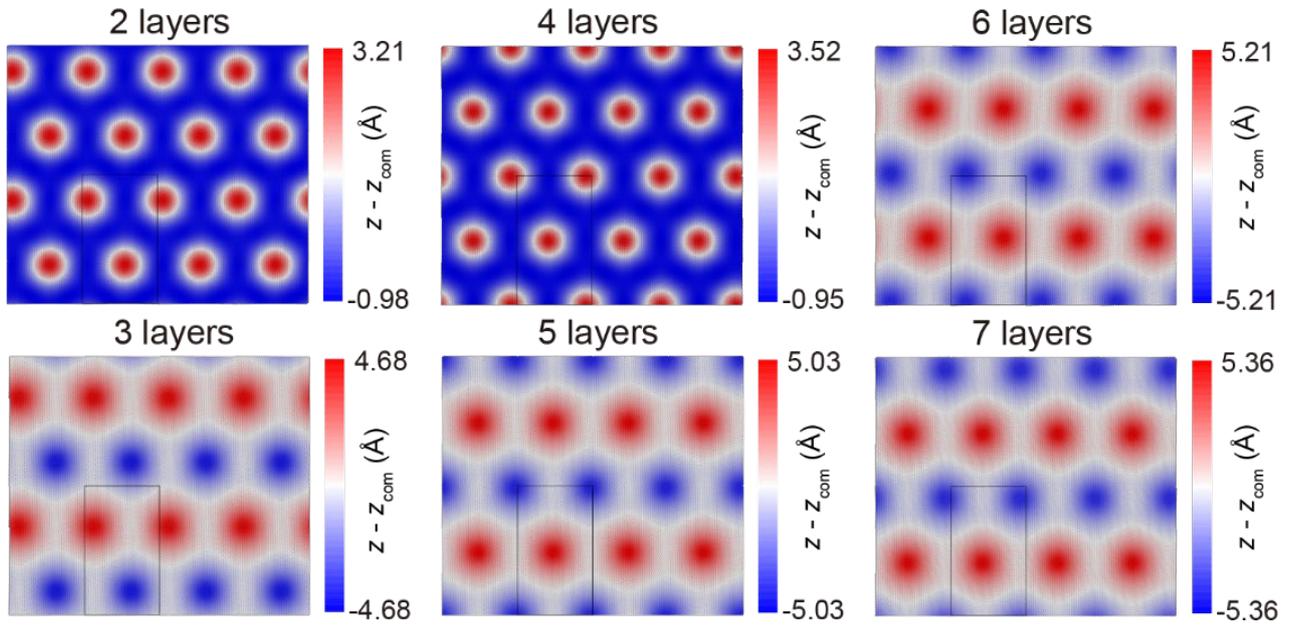

**Fig. 4. The out-of-plane deformation pattern of alternating twisted graphene heterostructures.** Each even numbered layer is rotated by 1.12° with respect to its adjacent odd numbered layers. To show the deformation pattern clearly, the simulation box (marked by black rectangle in each panel) is multiplied in the lateral directions. The color scales denote the perpendicular height of the top layer atoms with respect to its center-of-mass.

## Methods

**Model system.** To build the periodic alternating graphene/*h*-BN heterostructures while preserving the small experimental lattice mismatch (~1.8%) between graphene and *h*-BN, we followed the method outlined in Refs. [22,23] and built large rectangular supercells, where each graphene and *h*-BN layer contains 12,544 and 12,120 atoms, respectively. We considered two types of heterostructures with different periodic boundary conditions. One type, in which periodic boundary conditions are applied along the lateral directions alone is termed 2D PBC and the other type, where periodic boundary conditions are applied along the vertical direction as well is termed 3D PBC. Note that for the 3D PBC systems we tested two supercell thicknesses to verify the convergence of our results with respect to inter-image interactions. The Open Visualization Tool (Ovito) was used to visualize atomic configurations [47].

**Force fields.** Intralayer interactions for graphene and *h*-BN layers are modeled using the second generation REBO potential [35] and the Tersoff potential [36], respectively. Interlayer interactions are modeled using the registry-dependent interlayer potential ILP [37,38], which were recently reparametrized to better describe the sub-equilibrium regime for bilayer [24] and bulk configuration [39], respectively. The former and latter parametrizations are used for bilayer and multilayer graphene/*h*-BN heterostructures, respectively. We checked that using the latter parametrization for the bilayer configuration gives equivalent results, with a slightly reduced out-of-plane corrugation (8.54 Å vs. 8.6 Å).

**Simulation setup for energy minimization**. All geometry optimization simulations were performed with the LAMMPS simulation package [48]. The alternating graphene/*h*-BN heterostructures with various number of layers are first heated up to 1000 K and then cooled down to 300 K during 100,000 steps using the Nosé-Hoover thermostat with a time-step of 0.25 fs, followed by thermal equilibration at 300 K and zero pressure for another 100,000 steps with a time-step of 0.5 fs, where the pressure is controlled by the Nosé-Hoover barostat [49]. After annealing, the configurations are optimized by fixing the size of the simulation box using the FIRE algorithm [50] with a force convergence criterion of $10^{-5}$ eV/Å and a time-step of 1 fs. Once reaching the local minimum, the conjugate gradient (CG) method is used to relax the simulation box with a force convergence criterion of $10^{-3}$ eV/Å. To fully relax the system, the above procedure (FIRE + CG) is repeated for ten times and followed by a final optimization with FIRE algorithm with a force convergence criterion of $10^{-5}$ eV/Å (convergence tests for this procedure are provided in Sec. 2.2 of the Supporting Information). To verify that our results are not influenced by the lateral dimensions of the supercell we doubled the simulation box along the *x*-direction (see Sec. 2.1 of the Supporting Information) and repeated the optimization procedures. The results of these tests yielded the same optimized configurations thus confirming that our supercell is sufficiently large to describe studied superstructures.

We note that two other minimization protocols were considered (see details in Section 1 of the Supporting Information), where the size of the simulation box was kept fixed and the geometry optimization was performed with or without annealing. Comparing the results of the three protocols we conclude that cycles of annealing the system and relaxing the simulation box are important to avoid trapping in local minima (see Figs. S1-S2 in the Supporting Information) and that in-plane stresses play an important role for the resulting superstructures.

**Simulation Protocol for calculating Poisson's ratio and Young's modulus**. To calculate Poisson's ratio of the graphene/*h*-BN heterostructures, we applied strain ($\epsilon_x$ ranging from $-0.003$ to $0.003$) to the structures along the *x*-direction of the box (see Fig. 1) and the translational vectors ($L_y$) of the simulation cell in the *y*-direction was relaxed at zero pressure by a Nosé-Hoover barostat with a time constant of 1.0 ps for the lattice vectors relaxation [49]. The system temperature was controlled via a Nosé-Hoover thermostat, which was set to 5 K and 300 K in our simulations as detailed in the main text. The time step used in these simulations was set to 0.5 fs. The strain in the relaxed directions was then calculated using $\epsilon_y = (L_y - L_y^0)/L_y^0$, where $L_y^0$ is the equilibrium size of the simulations cell in the *y* direction at the desired temperature. Poisson's ratio is then calculated as $\mu_{xy} = -\epsilon_y/\epsilon_x$. By measuring the average stress ($\sigma_x$) along *x*-direction during the simulations, the Young's modulus of the alternating graphene/*h*-BN heterostructures was extracted from the linear regime of the $\sigma_x$ versus $\epsilon_x$ curve.

## Acknowledgments

W.O. acknowledges the financial support from the Planning and Budgeting Committee fellowship program for outstanding postdoctoral researchers from China and India in Israeli Universities and the support from the National Natural Science Foundation of China (Nos. 11890673 and 11890674). M.U. acknowledges the financial support of the Israel Science Foundation, Grant No. 1141/18 and the ISF-NSFC joint grant 3191/19. O.H. is grateful for the generous financial support of the Israel Science Foundation under grant no. 1586/17, Tel Aviv University Center for Nanoscience and Nanotechnology, and the Naomi Foundation for generous financial support via the 2017 Kadar Award.

## Author contributions

O.H. and M.U. directed the project. W.O. wrote the code and conducted the simulations. All authors conceived the original idea, designed the simulations, analyzed the results, and wrote the manuscript.

## Conflict of Interest

The authors declare no conflict of interest.

Supporting information for "**Parity-Dependent Moiré Superlattices in Graphene/*h*-BN Heterostructures: A Route to Mechanomutable Metamaterials**"

Wengen Ouyang, Oded Hod, and Michael Urbakh

*Department of Physical Chemistry, School of Chemistry and The Sackler Center for Computational Molecular and Materials Science, The Raymond and Beverly Sackler Faculty of Exact Sciences, Tel Aviv University, Tel Aviv 6997801, Israel.*

Email: odedhod@tauex.tau.ac.il

This Supporting Information contains the following sections:

1. Comparison of Minimization Protocols

2. Convergence Tests

3. Optimized Heterostructures for Stack Thicknesses Exceeding 9 Layers

4. Parity-Dependent Superlattices in Alternating Twisted *h*-BN Heterostructures

5. Raw Data for Calculating Young's Modulus and Poisson's Ratio

6. Thermal Expansion of Bulk Graphene/*h*-BN Heterostructures

7. Effect of Strain on the Out-of-plane Corrugation of Graphene/*h*-BN Heterostructures

# 1 Comparison of Minimization Protocols

## 1.1 *Effect of annealing and relaxing the simulation box*

To test the viability of the obtained optimized structure, three different energy minimization protocols were considered: (i) the structure was minimized without annealing and the size of the simulation box was kept fixed; (ii) the structure was annealed (first heating the system to 1000 K then gradually cooling it down to 300 K) before minimization and the size of the simulation box was kept fixed; and (iii) the structure was annealed as in (ii) and the size of the simulation box was relaxed during annealing and energy minimization processes. To demonstrate the importance of annealing and relaxing the simulation box during energy minimization, we first optimized the bilayer and trilayer graphene/*h*-BN heterostructures with the abovementioned protocols. For the bilayer case, protocols (i) and (ii) gave the same optimized structure with a well-defined moiré pattern and an out-of-plane corrugation (peak-to-deep value) of ~5.3 Å (see **Fig. S1**a), which is consistent with previous results [1,2]. Protocol (iii) resulted in a similar moiré pattern (see **Fig. S1**b), but with a considerably larger surface corrugation of ~8.6 Å. The reason is that protocols (i) and (ii) produce in-plane tensile stress (~0.15 GPa) that reduces the out-of-plane corrugation. Allowing for box relaxation the lateral dimensions contract by ~0.6%, resulting in a release of the in-plane tensile stress and a corresponding enhancement of the out-of-plane corrugation. The total energies of the optimized structures using protocols (i) and (ii) were -184,203 and -184,219 eV/supercell, respectively, for a supercell consisting of 49,288 atoms. For the trilayer graphene/*h*-BN heterostructure, the optimized structure from protocols (i) and (ii) exhibited hexagonal and trigonal superstructures with out-of-plane corrugation of ~0.2 Å and ~5.3 Å and total energies of -275,650 and -275,672 eV/supercell, respectively, for a supercell consisting of 36,744 atoms (a-b), which is double the size of the minimal supercell required to obtain the relevant superstructures. Notably, protocol (iii) resulted in the highly corrugated (~13 Å) flamingo pattern with a lower potential energy of -275,696 eV/supercell, as discussed in the main text (Figure 1c in the main text, also presented in **Fig. S2**c).

From these results it is clearly evident that the annealing procedure is important to identify local minima on the complex superstructure energy surface characterizing the studied systems. Furthermore, we find that in-plane stress can influence the obtained superstructure. Since we are interested herein in unstressed systems, we adopted protocol (iii) for all calculations presented in the main text. We note that future work on the effect of stress on the superstructure of layered heterostructures is being pursued.

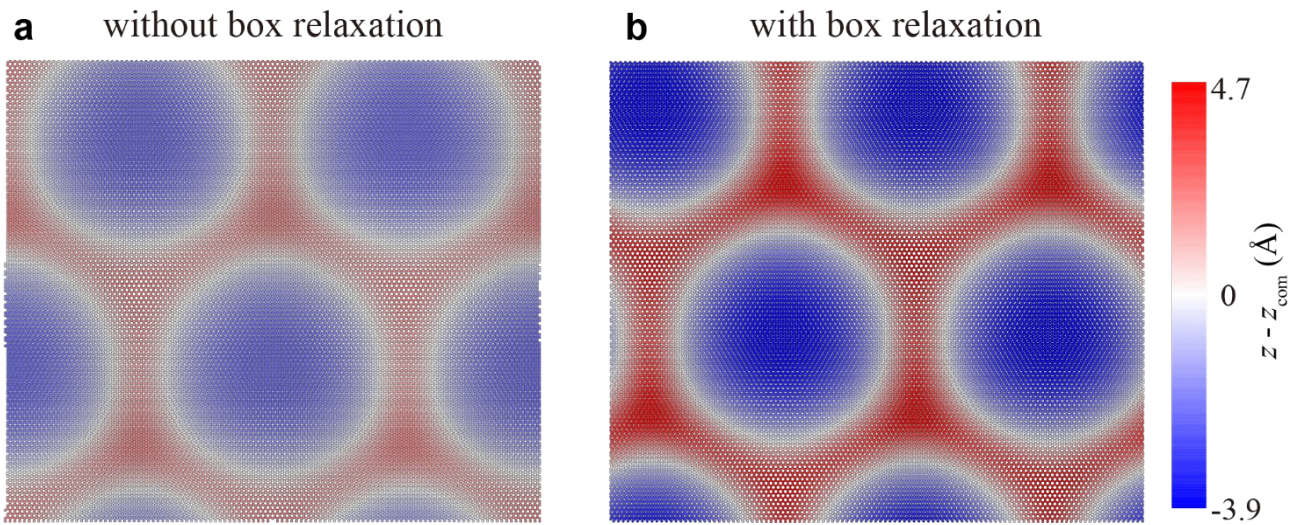

**Fig. S1. Effect of box relaxation on the energy minimization of a bilayer graphene/$h$-BN heterostructure.** Panels (a) and (b) correspond to the optimized structure without and with box relaxation, respectively. The color scale denotes the height of the top layer atoms with respect to its center-of-mass. To show the difference clearly, the same color scale is used in both panels. The full double-sized supercell that was used to perform these calculations is presented.

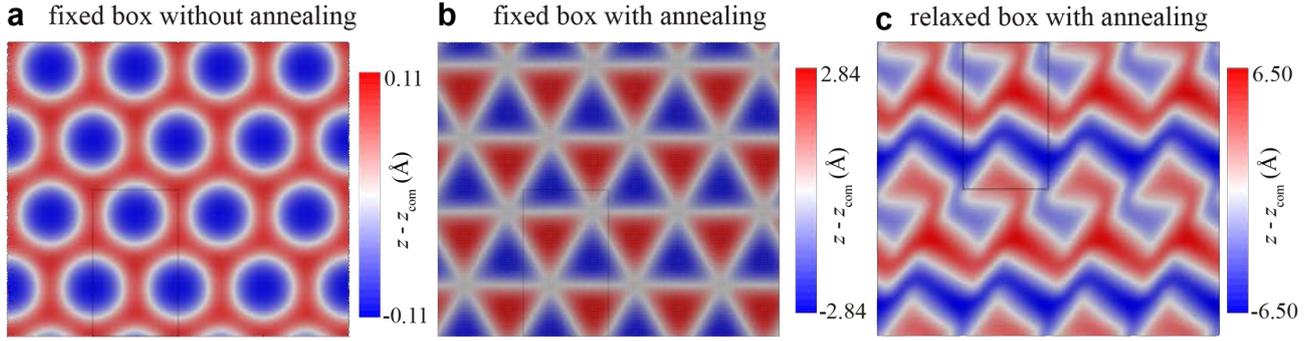

**Fig. S2. Effect of box relaxation and annealing on the energy minimization of a trilayer alternating graphene/*h*-BN heterostructure.** Panels (a) and (b) correspond to the fixed box dimensions optimized structure without and with annealing (protocols i and ii), respectively. Panel (c) shows the optimized structure with both annealing and box relaxation (protocol iii, same as Figurein the main text). The color scale denotes the height of the top layer atoms with respect to its center-of-mass. The modeled supercell is marked by the black rectangle in each panel.

*1.2 Relative Stability of the Optimized Trilayer Graphene/h-BN Superstructures*

To check the stability of the different trilayer graphene/*h*-BN heterostructures illustrated in **Fig. S2**, we performed nudged elastic band (NEB) simulations [3,4] to calculate the energy barrier to transform between them. Since the NEB method implemented in LAMMPS is implemented with fixed box size, we could only calculate the energy barrier between the states illustrated in **Fig. S2**a and **Fig. S2**b that have the same box dimensions. **Fig. S3** shows the results of the NEB calculations, where the total potential energy is plotted along the reaction coordinate ($r$) from the state with triangle pattern (**Fig. S2**b, $r = 0$) to the state with hexagonal pattern (**Fig. S2**a, $r = 1$). The fact that there is no barrier near $r = 1$ indicates that the hexagonal superstructure is either a saddle point or a very shallow local minimum on the potential energy surface (the estimated barrier is smaller than $8.62 \times 10^{-3}$ meV/supercell). The energy difference between the two states is ~21.7 eV/supercell, for a supercell size consisting of 36,744 atoms. We further note that there exists an equivalent mirror-image structure of the triangular pattern. The small barrier of ~0.43 eV/supercell (see **Fig. S3**b) separating them indicates that the two states can co-exist at room temperature. Finally, to evaluate the stability of the Flamingo superstructure we performed room temperature molecular dynamic simulations for the structure appearing in **Fig. S2**c, showing that the structure is stable over a simulation time of at least 200 ps (see Supplementary Movie 1) with no signatures of degradation.

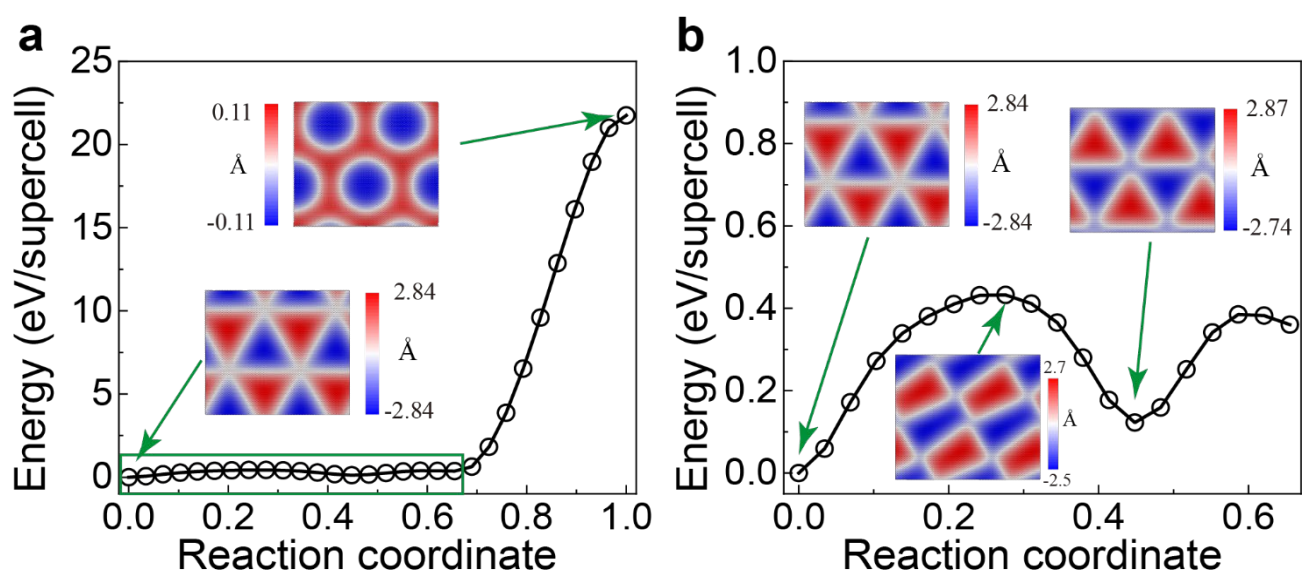

**Fig. S3. NEB calculations for a trilayer graphene/*h*-BN heterostructure with a rectangular supercell of 13.77 nm × 23.86 nm, which contains 36,744 atoms.** (a) Total potential energy along the reaction coordinate from the triangular state (b) to the hexagonal state (**Fig. S2**a). The lowest energy state is shifted to zero for clarity of the presentation. (b) Total potential energy along the reaction coordinate for the zoomed-in region marked by the green rectangle in panel (a) showing the transition pathway between two mirror-image triangular superstructures. Representative snapshots are chosen to show the local minima and transition state structures along the reaction coordinate. The color scale denotes the height of the top layer atoms with respect to its center-of-mass.

# 2  Convergence tests

## 2.1  Convergence with Respect to Supercell Lateral Dimensions

The supercell lateral dimensions used to obtain the results presented in the main text are 13.77 nm × 23.86 nm. To verify convergence with respect to these dimensions, we repeated our calculations for several stack thicknesses with a larger supercell size of 27.54 nm × 23.86 nm. As shown in **Fig. S4**, the deformation pattern of the optimized heterostructures with increased supercell size are the same as those presented in the main text, indicating that the supercell dimensions adopted in the main text are sufficiently large to obtain consistent results.

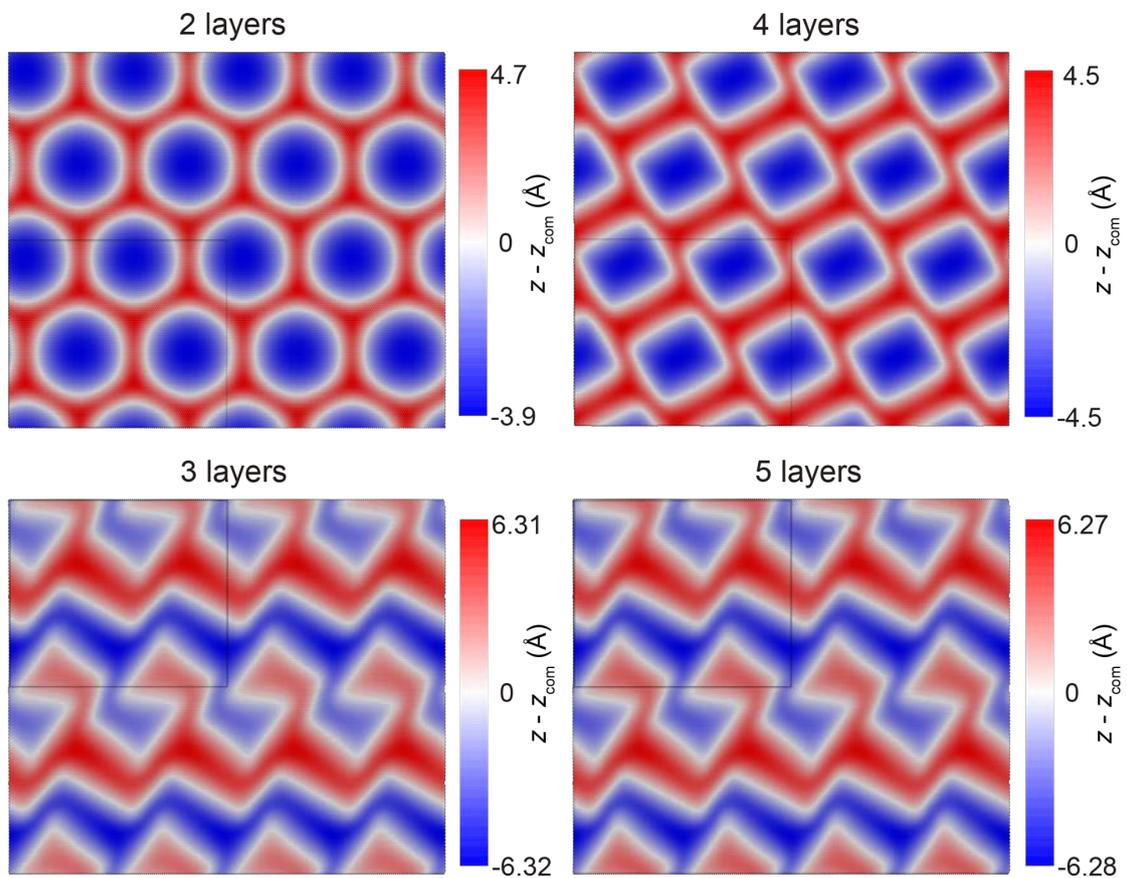

**Fig. S4. Size Effect.** Optimized structure for alternating graphene/*h*-BN heterostructures of lateral dimension of 27.54 nm × 23.86 nm and stack thicknesses ranging from 2 to 5 layers. The color scales denote the height of the top layer atoms with respect to its center-of-mass in each system. The modeled supercell is marked by the black rectangle in each panel.

## 2.2 Convergence with Respect to the Number of Optimization Cycles

As mentioned in the Methods section, the optimization procedure consisted of repeated cycles of geometry optimization with fixed simulation box dimensions using the FIRE algorithm followed by optimization of the supercell dimensions using the conjugate gradients method. To obtain the results presented in the main text, ten such cycles have been applied. In **Fig. S5** we present typical convergence curves of the potential energy of the system, the size of the simulation box, and the in-plane stress with respect to the number of optimization cycles. The energy, supercell dimensions, and in plane stress considered converge after 6 optimization cycles to within $7 \times 10^{-6}$ eV, $6 \times 10^{-6}$ nm, and $2 \times 10^{-6}$ GPa respectively, indicating that the results presented in the main text are well optimized.

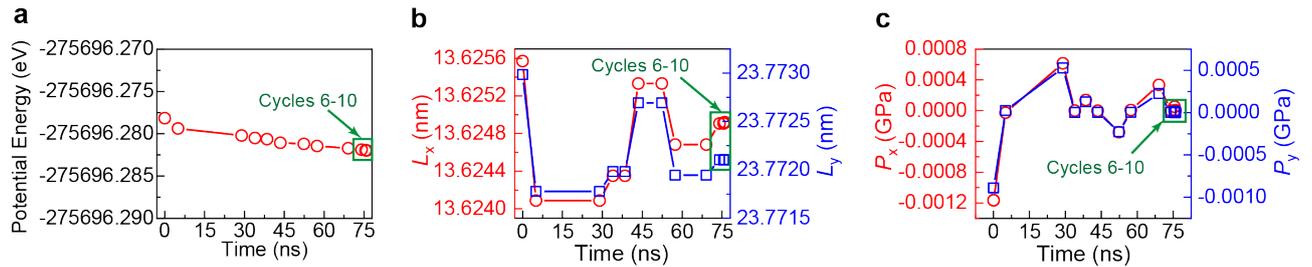

**Fig. S5. Convergence test** with respect to the number of optimization cycles for the trilayer graphene/*h*-BN heterostructure. Evolution of the (a) potential energy, (b) simulation box dimensions and (c) in-plane stress during the optimization cycles described in the Method section. The points inside the green rectangle belong to optimization cycles 6-10. The left and right axes in (b) and (c) represent the size of the simulation box and the in-plane stress of the system along *x* (open red cycles) and *y* (open blue rectangles) directions, respectively.

## 3  Optimized Heterostructures for Stack Thicknesses Exceeding 9 Layers

For brevity, we opted to present in the main text superstructures of alternating graphene/*h*-BN heterostructures with up to 9 layers. Here, we show results for thicker stacks of up to 16 layers. The results presented in **Fig. S6** further support our conclusion made in the main text that alternating graphene/*h*-BN heterostructures with more than 7 layers present the Flamingo superstructure with no parity effect.

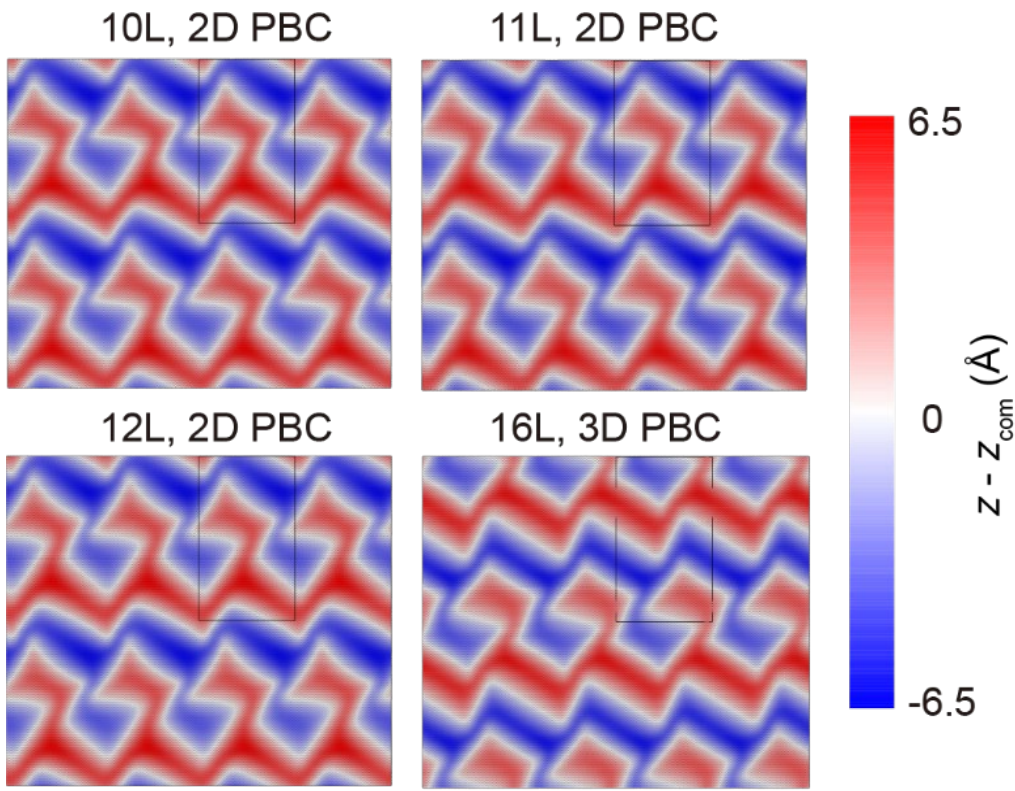

**Fig. S6. The deformation pattern for alternating graphene/*h*-BN heterostructures with number of layers ranging from 10 to 16.** To show the deformation pattern clearly, the simulation box (marked by black rectangle in each panel) is multiplicated along the lateral directions. The color scale denotes the atomic height with respect to the center-of-mass of the top layer.

# 4 Parity-Dependent Superlattices in Alternating Twisted *h*-BN Heterostructures

In the main text, we stated that the parity dependence of the out-of-plane deformation patterns observed for graphene/*h*-BN heterostructure is attributed to the interplay between the large-scale moiré superlattices formed at each interface in the heterostructures. Based on this we concluded that the parity dependence should be observed in other layered structures possessing large-scale moiré superlattices between adjacent layers and that such structures can be obtained also for homogeneous layered material stacks provided that adjacent layers are placed in a misaligned configuration. To support this conclusion, we presented in the main text results for few-layered alternating homogeneous graphene heterostructures. To further demonstrate the generality of our results we repeated our calculations for few-layered homogeneous alternating stacks of twisted *h*-BN, where every even numbered layer is rotated by 1.12° with respect to its adjacent odd numbered layers (the method for building laterally periodic twisted multilayer graphene and *h*-BN structures can be found in Ref. [5]). The optimized structures for 2-7 layered twisted *h*-BN stacks are presented in **Fig. S7**. The parity dependence is clearly observed for smaller number of layers and disappears above a thickness of 6 layers, consistent with the results obtained for alternating graphene/*h*-BN heterostructures and homogeneous twisted graphene stacks.

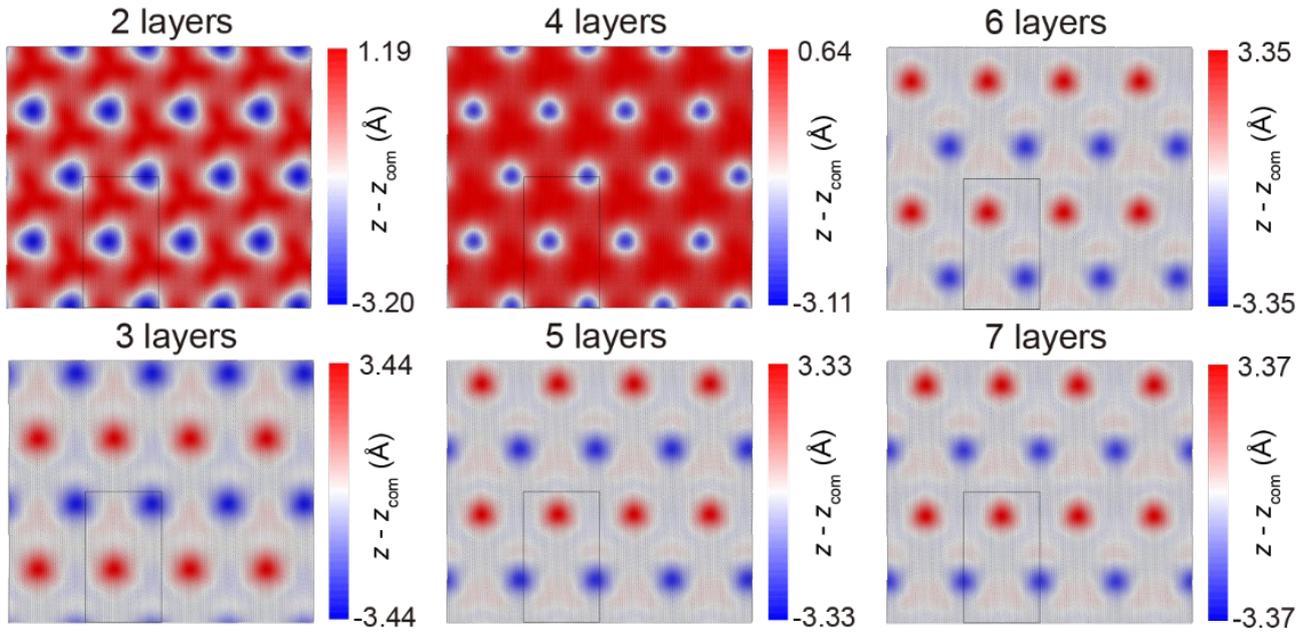

**Fig. S7. The out-of-plane deformation pattern for alternating twisted *h*-BN heterostructures.** Each even numbered layer is rotated by 1.12° with respect to its adjacent odd numbered layers. To show the deformation pattern clearly, the simulation box (marked by black rectangle in each panel) is multiplied in the lateral directions. The color scales denote the perpendicular height of the top layer atoms with respect to its center-of-mass.

## 5 Raw Data for Calculating Young's Modulus and Poisson's Ratio

In the main text we presented the calculated Young's modulus and Poisson's ratio for the various systems considered. For completeness, the raw data used to extract the values reported in the main text is provided herein for temperatures of 5 K and 300 K. **Fig. S8** and **Fig. S9** show the stress-strain relation along the $x$-direction for alternating graphene/$h$-BN heterostructures with the number of layers varying from 2 to 12 at 5 K and 300 K, respectively. The linear nature of these curves (linear fitting marked by the blue lines) indicates that the system is in the elastic regime and the slope of the stress-strain curve gives the Young's modulus of the heterostructures, which is listed in each panel of **Fig. S8** and **Fig. S9**. It is clear that the parity dependence of the Young's modulus on the number of layers is not affected by temperature although its value slightly changes. The larger error bars appearing in **Fig. S9** result from thermal fluctuations. Note that the heterostructures with 2 and 4 layers buckle at a compression strain of -0.2% at room temperature.

To calculate Poisson's ratio of the heterostructures, we applied strain along the $x$-direction and measured the corresponding variation of the simulation box size along the $y$-direction. **Fig. 10** and **Fig. S11** present the results obtained at 5 K and 300 K, respectively. As mentioned in the main text, the behavior of Poisson's ratio at room temperature is similar to that obtained at 5 K with a slight shift toward negative values, which may be attributed to the thermal expansion of the system with increasing temperature (see **Fig. S12**).

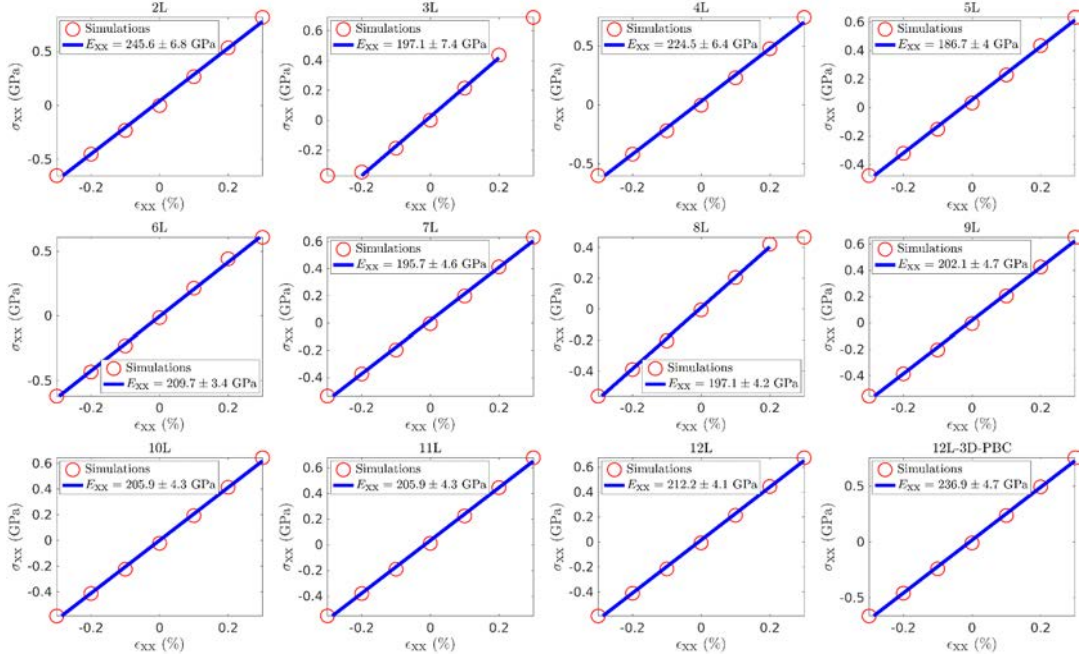

**Fig. S8. Stress-strain curves** obtained at a temperature of 5 K for alternating graphene/*h*-BN heterostructures with number of layers ranging from 2 to 12. The lower right panel provides results obtained using three-dimensional periodic boundary conditions.

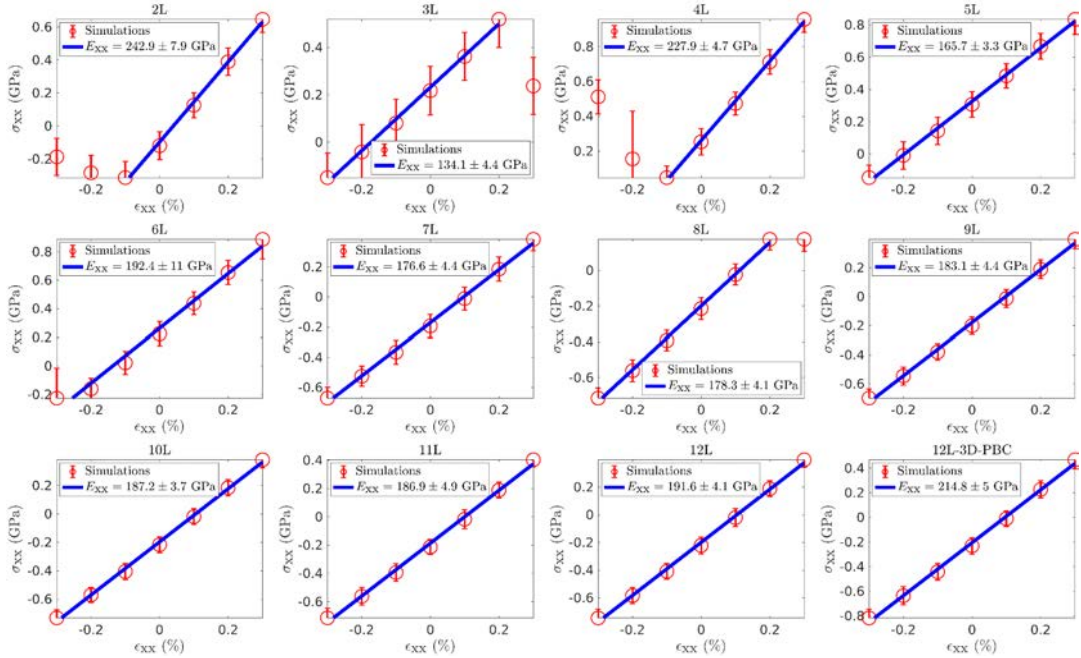

**Fig. S9. Stress-strain curves** obtained at a temperature of 300 K for alternating graphene/*h*-BN heterostructures with number of layers ranging from 2 to 12. The lower right panel provides results obtained using three-dimensional periodic boundary conditions.

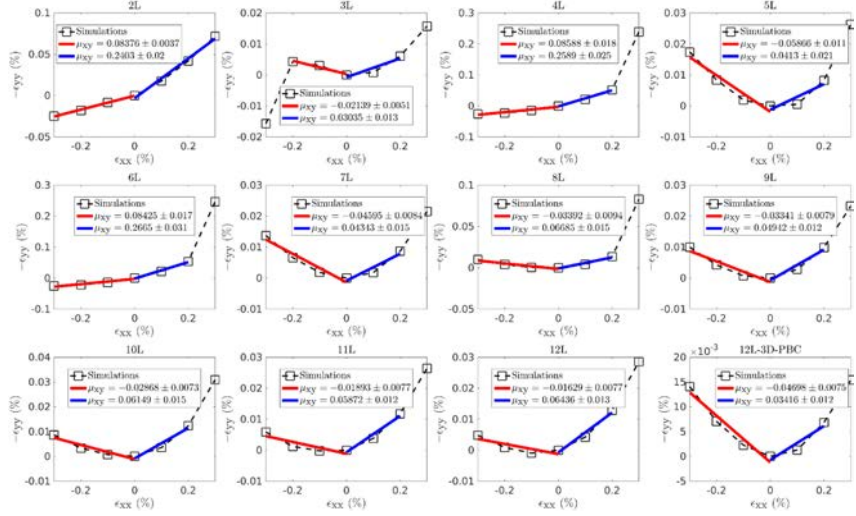

**Fig. 10. Strain along the *y*-direction ($\epsilon_{yy}$)** as a function of the applied strain along *x*-direction calculated at a temperature of 5 K for alternating graphene/*h*-BN heterostructures with the number of layers varying from 2 to 12. The lower right panel provides results obtained using three-dimensional periodic boundary conditions. The sign of $\epsilon_{yy}$ is inverted such that the sign of the calculated Poisson's ratio will match the slope of the curve.

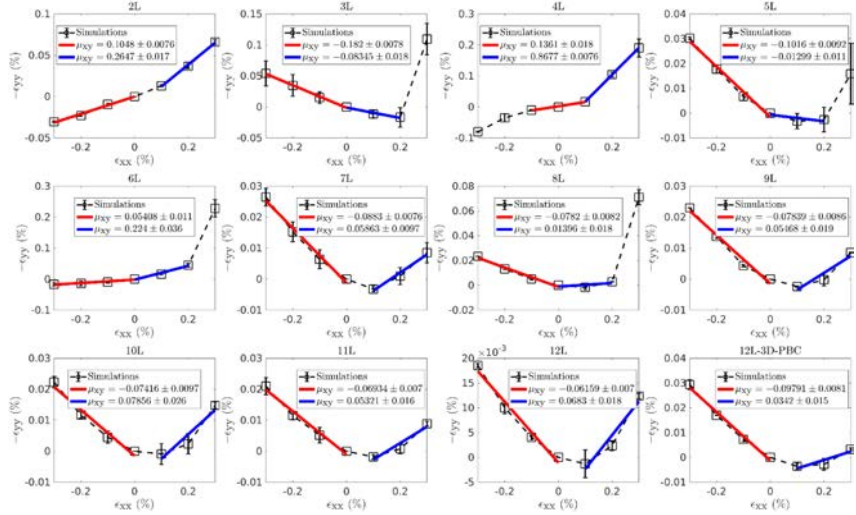

**Fig. S11. Strain along the *y*-direction ($\epsilon_{yy}$)** as a function of the applied strain along *x*-direction calculated at a temperature of 300 K for alternating graphene/*h*-BN heterostructures with the number of layers varying from 2 to 12. The lower right panel provides results obtained using three-dimensional periodic boundary conditions. The sign of $\epsilon_{yy}$ is inverted such that the sign of the calculated Poisson's ratio will match the slope of the curve.

## 6 Thermal Expansion of Bulk Graphene/*h*-BN Heterostructures

In this section we demonstrate the thermal expansion effect of alternating bulk graphene/*h*-BN heterostructures. To study the thermal effect along each direction, a bulk configuration (12 layers with three-dimensional periodic boundary conditions) is used, the system is equilibrated with an NPT ensemble at a desired temperature ($T_{\text{set}}$) for 200 ps. The thermal strain is then evaluated by averaging over the following 100 ps. As shown in **Fig. S12**, the thermal strain along the *x*, *y* (lateral) and *z* (vertical) directions (see Fig. 1 in the main text) increases linearly with temperature. The slope of the curve gives the thermal expansion coefficient, which is $3.2671 \times 10^{-6}$ K$^{-1}$, $2.5381 \times 10^{-6}$ K$^{-1}$ and $3.9237 \times 10^{-5}$ K$^{-1}$ along the *x*, *y* and *z* directions, respectively. From **Fig. S12** we can see that the thermal strain along the vertical direction (**Fig. S12**c) is an order of magnitude larger than that along the lateral directions (**Fig. S12**a-b), as expected for a van der Waals layered structure. Notably, the out-of-plane deformation corrugation remains nearly unaffected at the range of temperatures and simulation times considered (see **Fig. S12**d).

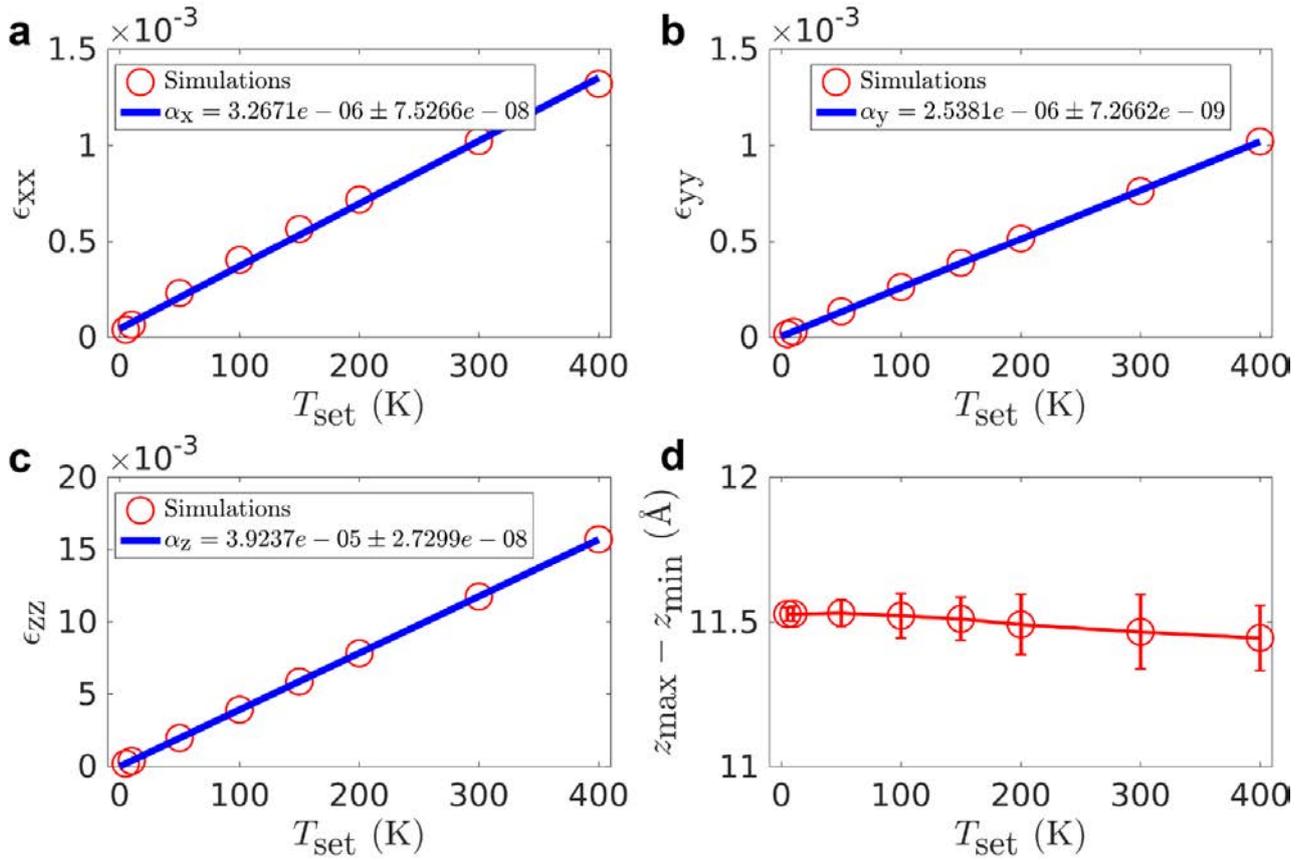

**Fig. S12. Thermal expansion of an alternating graphene/*h*-BN bulk heterojunction** (three-dimensional periodic boundary conditions with a 12 layers supercell). Panels **a**-**c** show the thermal strain along the lateral (**a**, **b**) and vertical (**c**) directions as a function of temperature (open red circles). The slope of the fitted curve (blue) gives the thermal expansion coefficient in the corresponding directions. Panel **d** shows the maximal (open red circles) and root mean square (open blue squares) out-of-plane corrugation as a function of temperature.

## 7 Effect of Strain on the Out-of-plane Corrugation of Graphene/*h*-BN Heterostructures

In the main text, we found that the values of Young's modulus for few-layered graphene/*h*-BN heterostructures are much smaller than that of monolayer graphene (~1000 GPa) [6] and *h*-BN (~865 GPa) [7]. We attributed this to the highly corrugated nature of the heterostructures that can flatten upon loading thus reducing in-plane covalent bond elongation effects. To support this point, we calculated the out-of-plane height for various number of layers of graphene/*h*-BN heterostructures as a function of applied strain. As can be seen from **Fig. S13**, the out-of-plane corrugation (peak-to-dip value) decreases as the applied stain increases, which shows a flattening effect during stretching, as suggested.

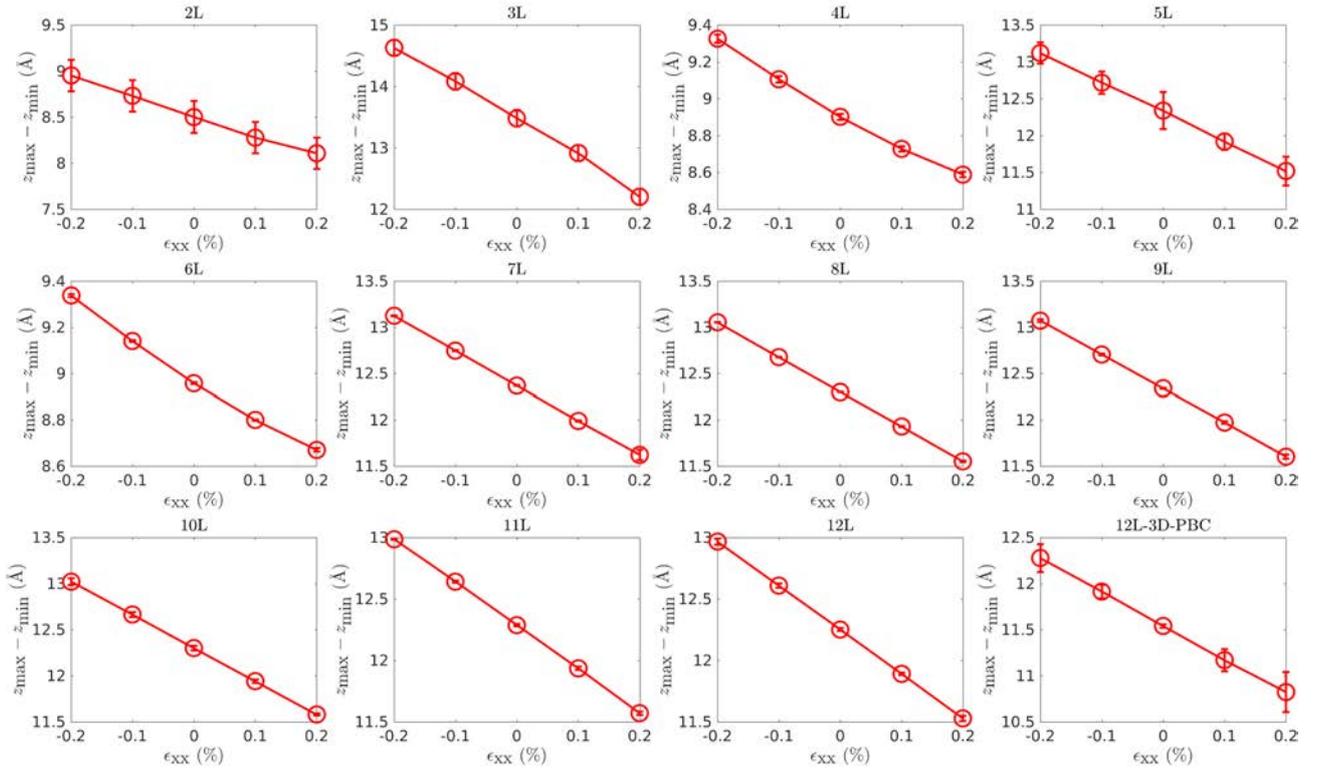

**Fig. S13. Out-of-plane corrugation (peak-to-dip value)** as a function of the applied strain along the *x*-direction calculated at a temperature of 5 K for alternating graphene/*h*-BN heterostructures with the number of layers varying from 2 to 12. Note that the out-of-plane corrugation is obtained by averaging the peak-to-dip values of all the layers in each heterostructures.